# Quantum enhanced electric field mapping within semiconductor devices

*D. Scheller[1]*, F. Hrunski[1]*, J. H. Schwarberg[2], W. Knolle[3], Ö. O. Soykal[4], P. Udvarhelyi[5], P. Narang[6,7], H. B. Weber[8], M. Hollendonner[1]†, R. Nagy[1]*

*[1]Insitute of Applied Quantum Technologies, Friedrich-Alexander-University Erlangen-Nürnberg, 91052 Erlangen, Germany*

*[2]Chair of Electron Devices, Friedrich-Alexander-University Erlangen-Nürnberg, 91058 Erlangen, Germany*

*[3]Leibniz Institute of Surface Engineering (IOM), 04318 Leipzig, Germany*

*[4]Photonic Inc., Coquitlam, British Columbia, V3K 6T1, Canada*

*[5]Department of Chemistry and Biochemistry, University of California, Los Angeles, CA 90095, USA*

*[6]Division of Physical Sciences, College of Letters and Science, University of California, Los Angeles, CA 90095, USA*

*[7]Department of Electrical and Computer Engineering, University of California, Los Angeles, CA 90095, USA*

*[8]Chair of Applied Physics, Friedrich-Alexander-University Erlangen-Nürnberg, 91058 Erlangen, Germany*

** These authors contributed equally*

*† Author to whom any correspondence should be addressed, maximilian.hollendonner@fau.de*

**Semiconductor components based on silicon carbide (SiC) are a key component for high-power electronics. Their behavior is determined by the interplay of charges and electric fields, which is typically described by modeling and simulations that are calibrated by nonlocal electric properties. So far, the 3D mapping of both the electric field and the concentrations of free charge carriers inside an electronic device remains a challenging task. To fulfill this information gap, we propose an operando method that utilizes single silicon vacancy ($V_{Si}$) centers in 4H-SiC. The $V_{Si}$ centers are at various positions in the intrinsic region of a *pin*-diode. To monitor the local static electric field, we perform Stark shift measurements based on photoluminescence excitation (PLE), which allows us to infer the expansion of the depletion zone and therefore to determine the local concentration of dopants. Besides this, we show that our measurements allow us to additionally obtain the local concentration of free charge carriers. The method presented here therefore paves the way for a new quantum-enhanced electronic device technology, capable of mapping the interplay of mobile charges and electric fields in a working semiconductor device with nanometer precision.**

The performance of semiconductor devices is crucially determined by the complex interplay between charge carriers and electric fields. Models that balance microscopic quantities, like for instance the formation of depletion zones or the distribution of free charge carriers, have allowed to obtain an in-depth understanding of the underlying physical processes. Their validity was so far mainly confirmed by comparing, for example, the resulting capacitance or current [1] as a function of the applied voltage, as electric fields and charge carrier distributions inside the semiconductor are hardly experimentally accessible [2,3]. Electric fields originating from crystal defects can be mapped using the edge transient-current technique [4,5] but determining the local doping concentration remains challenging. Nevertheless, these models paved the way for one of the greatest technological advances in history: semiconductor electronics.

The described interplay takes mainly place in the depletion zones between *p* and *n* doped regions. An externally applied voltage on a *p* or *n* doped region leads to local changes in the band structure [see Fig. 1 (a)], which is related to the spatial electric field $E$ across the junction as

$$E = \frac{1}{e} \nabla \varepsilon_V(r). \quad (1)$$

Here $\varepsilon_V(r)$ is the energy of the valence band maximum and $e$ the electron charge. It is therefore highly desirable to measure the electric field $E$ with a high spatial resolution to gain a clear understanding of the device behavior.

It is only now that state-of-the-art quantum technology provides a way to microscopically investigate the interplay of electric fields and charges in a semiconductor component. For instance, widefield NV magnetometry allows to measure currents inside a photovoltaic cell, attached to a layer of NV centers [6]. A color center, placed inside a device, can be used as a highly sensitive and robust quantum sensor to map electric fields [7,8] and charge densities. Such a quantum sensor offers the unique possibility to directly image an electric field and provides a quantitative estimation of charge carriers inside a device and – even more sophisticated - during operation. These insights are currently mainly accessed by numerical simulations. Obtaining them experimentally may deliver more detailed information about the critical influence of imperfections in real devices and hence has the potential to shift what is technologically possible.

For our experiments we use 4H-SiC, which is a well-established material for high-power semiconductor electronic applications [9]. In addition, 4H-SiC is also an excellent host material for quantum technology applications [10–12]. Especially color centers inside the crystal lattice show excellent optical and spin properties [13–17], establishing them as an ideal platform for quantum communication [18,19], quantum computing [20–22], and quantum sensing [23–26]. Among the different color centers in 4H-SiC, the silicon vacancy ($V_{Si}$) center stands out due to significantly reduced spectral diffusion [27], the integrability into nanophotonic devices [28], and long spin coherence times [29,30]. This makes the $V_{Si}$ center a very promising candidate for quantum technology applications. A further advantage is the capability to integrate $V_{Si}$ centers into semiconductor structures [25,31], which opens the unique possibility to monitor the interplay between electric fields and free charge carriers with nanometer resolution at various positions in an electronic device.

We show here the integration of single $V_{Si}$ centers into a *pin*-diode [see Fig. 1 (b)] for in-situ and even operando mapping of electric field measurements and determination of the charge carrier concentrations. For the measurements reported here spectroscopy of the $V_{Si}$ center's narrow optical resonances between ground and excited state [see Fig. 1 (c,d)] is performed and we find frequency shifts of their positions due to the Stark effect. Besides this, we show that the voltage-dependent onset of the Stark shift in correlation with the spatial position of a $V_{Si}$ center enables spatially resolved monitoring of the formation of an electric field across the *pin*-diode [see Fig. 1 (a)] and therefore, the local charge carrier information. Our results mark an essential step towards an improved understanding of a semiconductor device, as the method presented here can be used at arbitrary locations inside the material and allows to infer the physical parameters of the device with nanometer-resolution. Furthermore, we show that the voltage-dependent onset of the Stark shift is also present in optically detected magnetic resonance (ODMR) measurements. As ODMR is applicable at room temperature, the method presented here paves the way for a new quantum-enhanced electric field mapping of semiconductor devices.

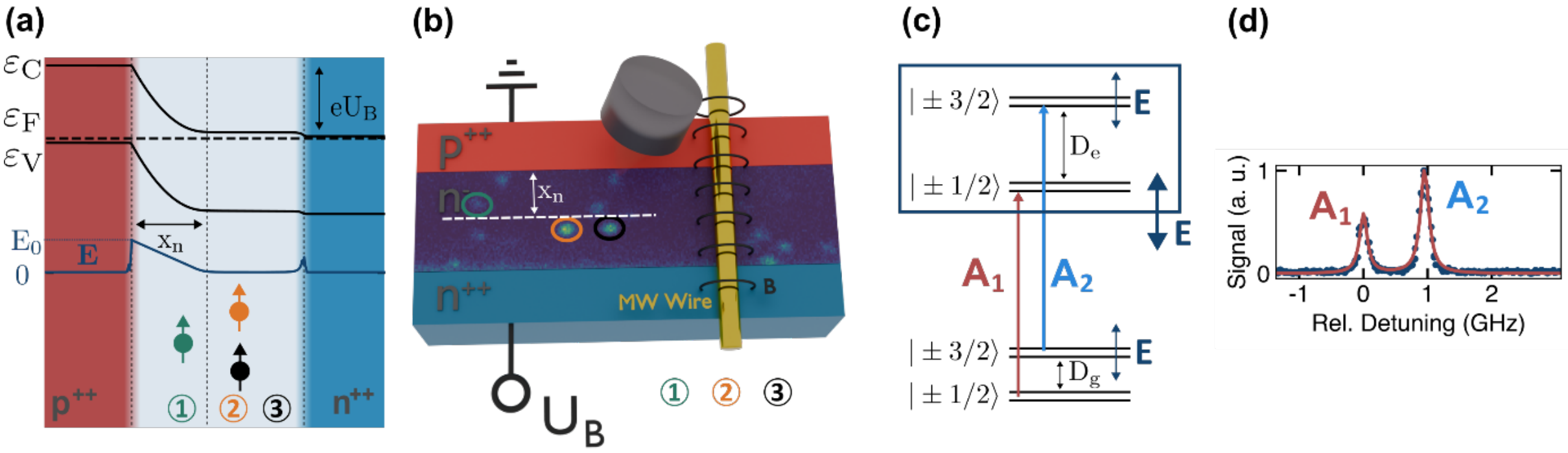

***Figure 1:*** *(a) The properties of a semiconductor device, including the electric field $E$, are determined by its band diagram and can be extracted via operando measurements at various points inside the low n-type doped layer (further referred to as "intrinsic layer" of the pin-diode). At the interface between p++ and the intrinsic layer, a depletion zone of width $x_n$ forms, which can be determined via measurements of various $V_{Si}$ centers. (b) Schematic of our setup, composed of a 4H-SiC pin-diode, a microwave (MW) wire for spin manipulation, and a high-NA objective for excitation and collection of the emission. Color centers at a depth of $\approx 5 \pm 1\,\mu m$ from the surface and different lateral positions inside the intrinsic region allow characterization of the semiconductor device, e.g., the extent of the depletion zone $x_n$. (c) Without a magnetic field, the $|\pm 1/2\rangle$ and $|\pm 3/2\rangle$ states are degenerate and connected to their excited states through $A_1$ and $A_2$ transitions. Axial electric fields influence both the ground (excited) state zero-field splitting $D_g$ ($D_e$) as well as the zero-phonon-line between the ground and excited state with spin-dependent resonances $A_1$ and $A_2$. (d) Both $A_1$ and $A_2$ transitions are identified via Photoluminescence-Excitation (PLE) measurements.*

The used *pin*-diode's $n^{++}$ and $p^{++}$ regions contain dopings in the range of $1 \times 10^{19}\ cm^{-3}$ and are separated by a low *n*-type doped area of width $4.1\ \mu m$, which we denote in the following as the intrinsic region of the diode. Based on our Stark shift measurements we find a concentration of dopants of around $9 \times 10^{14}\ cm^{-3}$ (see discussion below), which deviates from the value specified by the manufacturer of $2 \times 10^{14}\ cm^{-3}$. The single $V_{Si}$ centers within the sample are created via electron irradiation followed by an annealing step. Further information about the sample can be found in the Supplementary Information [32] (SI1). Our measurements are carried out at room temperature and 5 K. Optical measurements are performed at 5 K with a homebuilt confocal microscope, where the optical resolution is diffraction limited at $> 500\ nm$. The sample is excited from a cleaved edge of the sample through a high-NA objective with a 730 nm laser for off-resonant and a tunable diode laser at 916 nm for resonant excitation. The emitted photons are filtered with a 950 nm long pass before being transmitted to a single photon detector with a quantum efficiency above *97 %* for the emission spectrum of the $V_{Si}$ center. The manipulation of the spin states is realized by an arbitrary waveform generator and a microwave wire on top of the sample. Detailed information about the home-built confocal setup including Refs. [14,33–52] is shown in the Supplementary Information [32] SI2. For our measurements, we use the *k*-$V_{Si}$ center (*$V_2$*). The $V_{Si}$ center possesses a spin $S = \frac{3}{2}$ with two spin-conserving optical transitions, denoted as $A_1$ and $A_2$ [see Fig. 1(c)]. These two resonances can be detected by optical spectroscopy of the $V_{Si}$ center's fine structure, a method denoted as photoluminescence excitation (PLE) [Fig. 1 (d)]. The energy of both optical transitions depends mainly on axial electric fields, which can be taken into account via the Stark shift Hamiltonian [47,51]:

$$H_{gs} = (D + d_z E_z) S_z^2. \tag{2}$$

Here $2D = 70\ MHz$ is the zero-field splitting (ZFS) of the *k*-$V_{Si}$, which is shifted by an axial electric field $E_z$. $S_z$ is the axial spin-3/2 operator in the reference frame of the $V_{Si}$ center, where the z-axis is defined by the c-axis of the 4H-SiC lattice. Even though this Hamiltonian was derived for NV-centers in diamond, we do not expect changes for axial electric fields due to the $C_{3v}$ symmetry of both NV and $V_{Si}$ centers. In Eq. (2) contributions of transverse electric fields can be neglected, as the electric field is aligned with the c-axis of the crystal due to the device geometry, i.e. $E \parallel c$.

Eq. (2) indicates that the axial electric field induced by the *pin*-diode leads to a Stark shift of the optical transitions $A_1$ and $A_2$ which can be measured through PLE at cryogenic temperatures (5 K). An FEM-simulation of the electric field under the Lorentz local field approximation in dependence of the applied voltage is shown in Fig. 2 (a) (upper graph, further information about the FEM-simulations can be found in the Supplementary Information [32] (SI3)). The electric field is generated by operating the *pin*-diode in reverse bias mode, where a positive voltage is applied on the *n*$^{++}$-doped area while the *p*$^{++}$-layer is grounded. The local electric field $E$ at the position of the $V_{Si}$ centers within the intrinsic layer increases with the applied reverse bias voltage. In the lower graph [Fig. 2(a)], three $V_{Si}$ centers are color-coded at different positions in the intrinsic region (1.61 $\mu m$ and 2.71 $\mu m$ from the *p*$^{++}$-layer) where the local electric field reaches values up to 35.43 $\frac{MV}{m}$ for $V_{Si1}$ (28.04 $\frac{MV}{m}$ for $V_{Si2,3}$) at 30V. These $V_{Si}$ centers are also shown with the same colors in the confocal map in Fig. 1 (b).

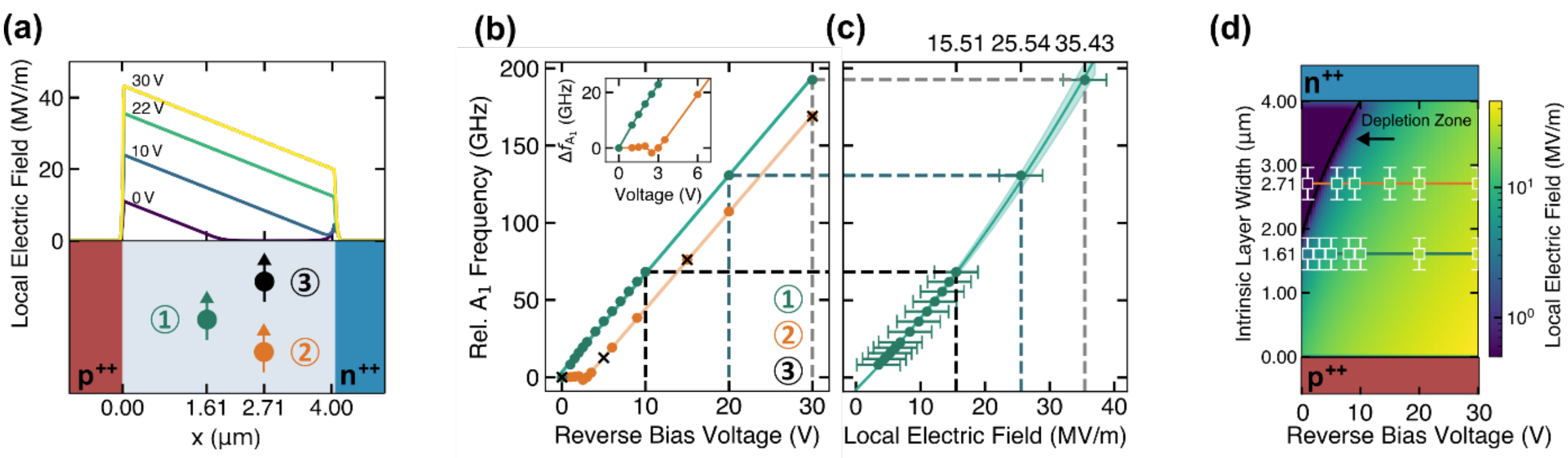

***Figure 2:*** *2D electric field mapping. (a) Local electric field, simulated across the intrinsic layer for voltages from 0 V to 30 V, including the Lorentz Local Field Approximation (see SI3 [32]). (b) Measured shift of $A_1$ lines upon different applied voltages. Inset: The relative shift of the $A_1$ frequency ($\Delta f_{A_1}$) of $V_{Si\,2}$ starts at around 2.6 V. (c) Stark shift of $V_{Si\,1}$ as a function of the local electric field, incorporating electric field uncertainty due to uncertainty of the $V_{Si}$ position (horizontal error bars). The shaded region displays the fit with 1-sigma uncertainty. Measurements of the PLE frequency in (b) allow for the reconstruction of the electric field values. (d) Simulation of the electric field distribution in the intrinsic layer, indicating that for 0 V $V_{Si\,1}$ (green) is inside and $V_{Si\,2}$ (orange) outside the depletion zone (black line). Dopant concentrations as used in the simulations entering into Fig. (c) and (d) is $9 \times 10^{14}\ cm^{-3}$. Vertical error bars in (b) and (c) are smaller than the data point.*

We perform PLE measurements to determine the Stark shift caused by the electric field $E$ within the intrinsic region. As can be seen from Fig. 2 (b), the $A_1$ transitions are shifted with respect to the applied voltage and we find that the relative distance between the optical transitions of $A_1$ and $A_2$ stays constant at $\approx 1\ GHz$ (see Supplementary Information [32] SI5). The Stark shift of $V_{Si\,1}$ center (marked green in Fig. 2 (a)) increases linearly, but for $V_{Si\,2,3}$ centers (orange and black color code) the PLE frequency is only shifted for reverse bias voltages larger than $\approx 2.6\ V$ [Inset Fig. 2 (b)]. This behavior can be understood by considering the depletion from the *p*$^{++}$ into the intrinsic layer, which scales as $x_n(V) \propto \sqrt{(V + V_{bi})/N_D}$, in terms of the built-in voltage $V_{bi} = 2.95\ V$ and $N_D$ the concentration of dopants in the intrinsic layer. As $V_{Si\,2}$ is located 2.71 $\mu m$ away from the *p*$^{++}$-layer [Fig. 2 (a)], the voltage-dependent extension of the depletion region can be used to infer the concentration of dopants in the intrinsic layer, which we find to be $N_D = (8.7 \pm 2.2) \times 10^{14}\ cm^{-3}$ (see SI4 [32] for the detailed calculation), resulting in a depletion region of $x_n = 1.9\ \mu m$ for zero applied reverse bias voltage. This value is in good agreement with capacitance-voltage (CV) measurements (see SI1 [32]), yielding $N_D = 8.4 \times 10^{14}\ cm^{-3}$. By simulating the electric field [Fig. 2 (a,c)], the applied voltage can be converted into the resulting local electric field for $V_{Si\ 1,2}$. The PLE shift is fitted with

$$\Delta f_{A_1} \propto -d \cdot E_z - \frac{\alpha}{2} E_z^2, \tag{3}$$

which allows to extract the dipole moment $d$ and the polarizability $\alpha$ of $V_{Si\ 1}$ as, respectively, $|d| = 4.33 \pm 0.15\ \frac{GHz}{\left(\frac{MV}{m}\right)}$ and $\alpha = -0.09 \pm 0.01\ \frac{GHz}{\left(\frac{MV}{m}\right)^2}$ [Fig. 2 (c)]. These values are in good agreement with the previously measured dipole moment of the *k*-$V_{Si}$ of $d = 3.65\ \frac{GHz}{\left(\frac{MV}{m}\right)}$ [53] and are comparable to first-principles estimates of the *k*-$V_{Si}$'s polarizability [37] and experimental measurements of hexagonal silicon vacancy (*h*-$V_{Si}$) centers [54]. A detailed discussion about the derived dipole moment and the polarizability can be found in SI6 [32]. Based on first-principles calculations, we determine $d = 1.83 \pm 0.01\ \frac{GHz}{\left(\frac{MV}{m}\right)}$ and $\alpha = -0.0254 \pm 0.0006\ \frac{GHz}{\left(\frac{MV}{m}\right)^2}$ (see SI7 [32] for detailed information). The differences from our measured values are probably related to the functional used in the first-principles calculation and to piezo effects.

The derived values of the dipole moment and polarizability are inherent properties of the $V_{Si}$ center and are independent of the 4H SiC device. For this reason, Stark Shift measurements of $V_{Si}$ centers as a function of the applied reverse bias voltage allow for the reconstruction of the local electric field of the color center under consideration. This reconstruction is exemplarily performed in Fig. 2(b,c) for voltages of 10 V, 20 V and 30 V, resulting in electric fields of, respectively, 15.03$MV/m$, 24.76 $MV/m$ and 34.34 $MV/m$ at the position of $V_{Si1}$. The presented measurement therefore opens the possibility to compare the measured electric fields against simulations of a semiconductor device [see data points in Fig. 2 (d)]. The sensitivity of the measurements presented in Fig. 2 is 18 $\frac{kV/m}{\sqrt{Hz}}$ (see SI8[31]), which is comparable to single NV-centers [47] in bulk diamond. As the electric break-down field of typical SiC devices is on the order of 328 $MV/m$ [55], $V_{Si}$ centers are therefore excellent quantum systems to infer the mechanisms of the device failure, e.g. the electric-field dependent onset of impact ionization [56].

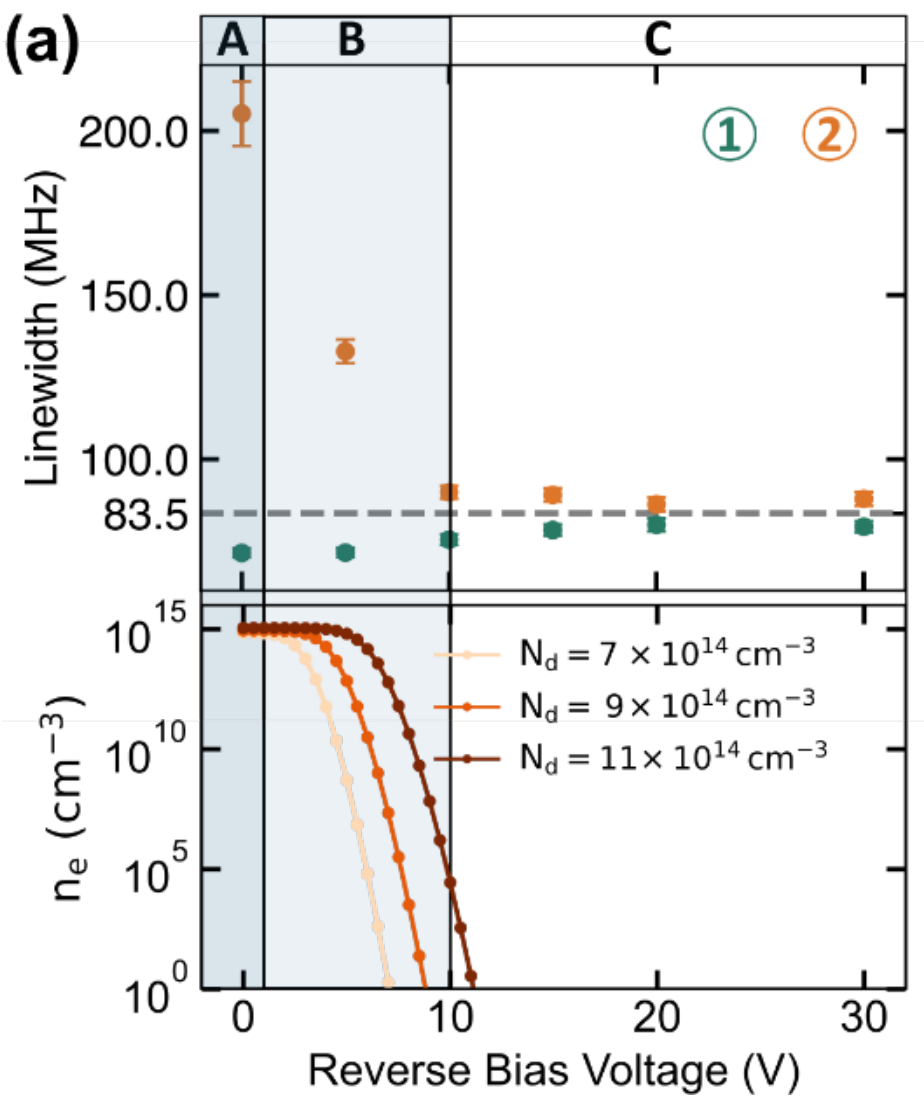

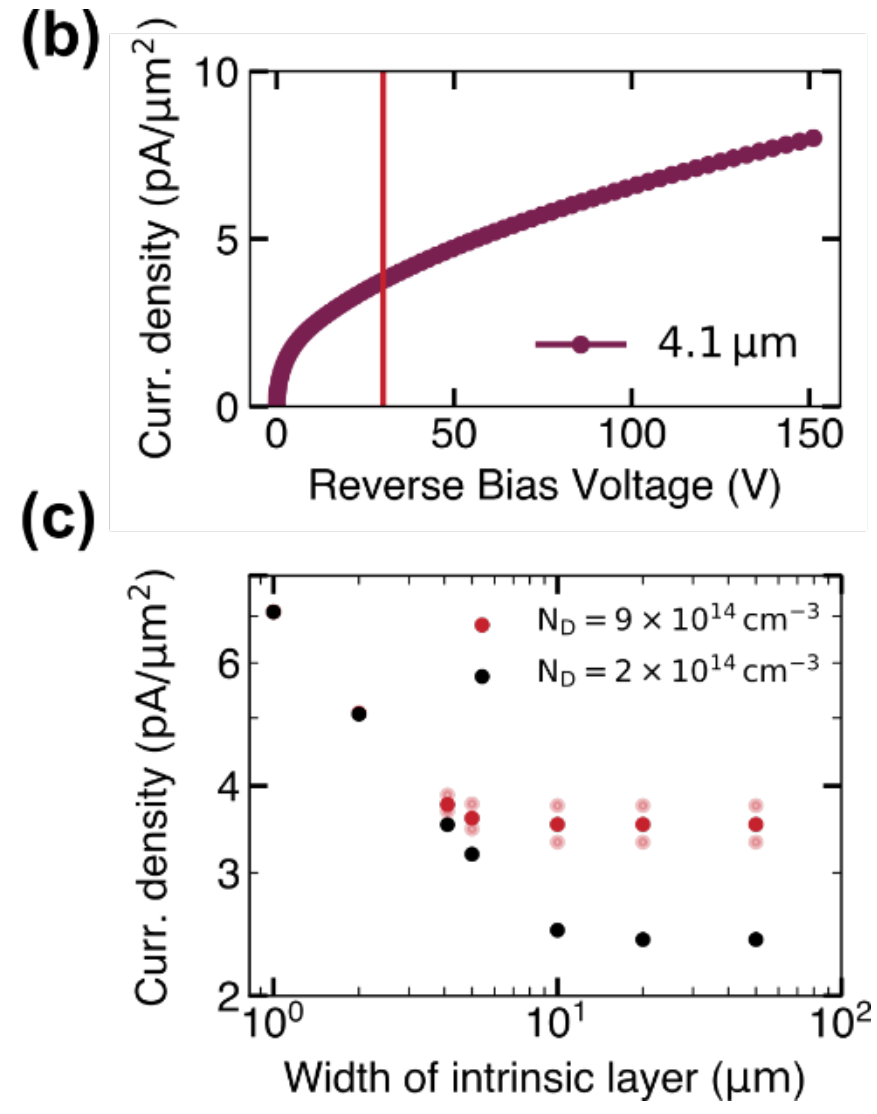

***Figure 3:*** *(a) Dependency of the linewidth on the position of the vacancy inside the intrinsic layer and the applied voltage. Upper panel: Linewidth of* $V_{Si\ 1,2}$ *centers as a function of applied reverse bias voltage. Lower panel: Simulated concentration of electrons at the positions of* $V_{Si\ 1,2}$ *centers. (b) Simulation of the reverse current density of the pin-diode considered here (width* $4.1\ \mu m, N_D = 9 \times 10^{14}\ cm^{-3}$*) at different reverse bias voltages. (c) Simulation of the reverse current density at 30 V as a function of the width of the intrinsic layer for* $9 \times 10^{14}\ cm^{-3}$ *(red) incorporating the uncertainty of the doping concentration (light red) and* $2 \times 10^{14}\ cm^{-3}$ *(black).*

Whereas the exact position of the PLE lines depends on the local electric field $E$ seen by the single $V_{Si}$ center, their optical linewidth is the result of charge fluctuations. Free charges can either stem from ionized defects, such as nearby ionized carbon vacancies [31], or incorporated dopants [57] residing in the vicinity of the $V_{Si}$ center. The resulting electric field $E_{fluc}$ created by the fluctuating charges couples to the permanent electric dipole moment $d$ in the ground and excited state of the $V_{Si}$ as $H \sim d \cdot E_{fluc}$. Therefore, the optical linewidth depends on the fluctuating electric field around the $V_{Si}$ and a reduction of mobile charges in the local environment of the $V_{Si}$ center causes a reduced optical linewidth of both $A_1$ and $A_2$. This can be achieved within a *pin*-diode due to the formation of a depletion region by applying reverse bias voltages [46]. We find that the optical linewidth strongly depends on the position of the color centers and can range from $205.2 \pm 9.8\ MHz$ to $71.5 \pm 1.4\ MHz$ in the absence of any applied reverse bias voltage, as shown in Fig. 3(a). We attribute this behavior to the width $x_n$ of the depletion region in the absence of an applied reverse bias voltage. As the sample's built-in voltage of $V_{bi} = 2.95\ V$ causes the depletion zone to extend $1.9\ \mu m$ into the intrinsic layer, $V_{Si\ 1}$ ($x = 1.6\ \mu m$, see Fig. 2 (a)), resides in the space-charge region of the diode, contrary to $V_{Si\ 2}$ ($x = 2.7\ \mu m$). This indicates that – unlike $V_{Si\ 2}$ – the charge environment around $V_{Si\ 1}$ center is initially already depleted. An increase of the reverse bias voltage of the *pin*-diode shifts the depletion region further into the intrinsic area where eventually also $V_{Si\ 2}$ center is in the space charge region. As a consequence, the concentration of surrounding charges around $V_{Si\ 2}$ is significantly reduced and the optical linewidth starts to narrow. If the reverse bias voltage is increased $\geq 10\ V$, the optical linewidth of $V_{Si\ 2}$ center reaches its plateau of $\approx 80\ MHz$. This finding stays in good agreement with the charge depletion simulation shown in Fig. 3 (a) (bottom), indicating that free charges within the intrinsic region are fully depleted at a reverse bias voltage of $\geq 10\ V$.

Whereas the linewidth of $V_{Si\ 2}$ center decreases with increasing reverse bias voltage, the linewidth of $V_{Si\ 1}$ center gets broadened [see Fig. 3 (a) region B], which cannot be explained by the depletion of free charges within the intrinsic area. Based on FEM simulations of the *pin*-diode, it can be shown that with increasing applied reverse bias, a reverse current starts to flow [see Fig. 3 (b)]. This behavior is in accordance with IV measurements of our pin-diode (see SI1 [32]). This reverse bias current causes a rise of mobile charge carriers flowing in the vicinity of $V_{Si\ 1}$ center and therefore increases its optical linewidth. As the concentration of charges due to this current is on the order of $\sim 10^8\ cm^{-3}$ (see Supplementary Information [32] SI9), the broadening effect is small compared to the narrowing effect of the formation of the space charge region (which reduces the charge concentration by $\sim 10^{15}\ cm^{-3}$). It can be seen from Fig. 3 (a) (region C) that $V_{Si\ 1,2}$ centers both approach a linewidth of around 80 MHz for high voltages. Here both $V_{Si}$ are in the space charge region and the same current flows nearby. This linewidth is larger than the $V_{Si}$ center's $A_2$ optical lifetime limit of 14 MHz [48], which we attribute to the reverse current becoming the dominant mechanism at higher voltages. It should be emphasized, that the effective reverse bias current might deviate from Fig. 3(b) due to imperfections of the studied *pin*-diode (see SI1 [32] for additional discussion). As discussed in SI10 [32], we do not expect magnetic fields originating from reverse currents to influence the PLE measurements reported in this manuscript. Simulations show that the reverse current can be minimized by using an intrinsic layer with a reduced concentration of dopants as well as an intrinsic region with a width of $\geq 10\ \mu m$ [Fig. 3 (c)]. The PLE linewidth can additionally be reduced with an improved epitaxial layer quality, which reduces deep traps and thus prevents Shockley-Read-Hall generation [58] of charges and a structured $p^{++}$-layer minimizing edge currents (see discussion in SI1 [32]).

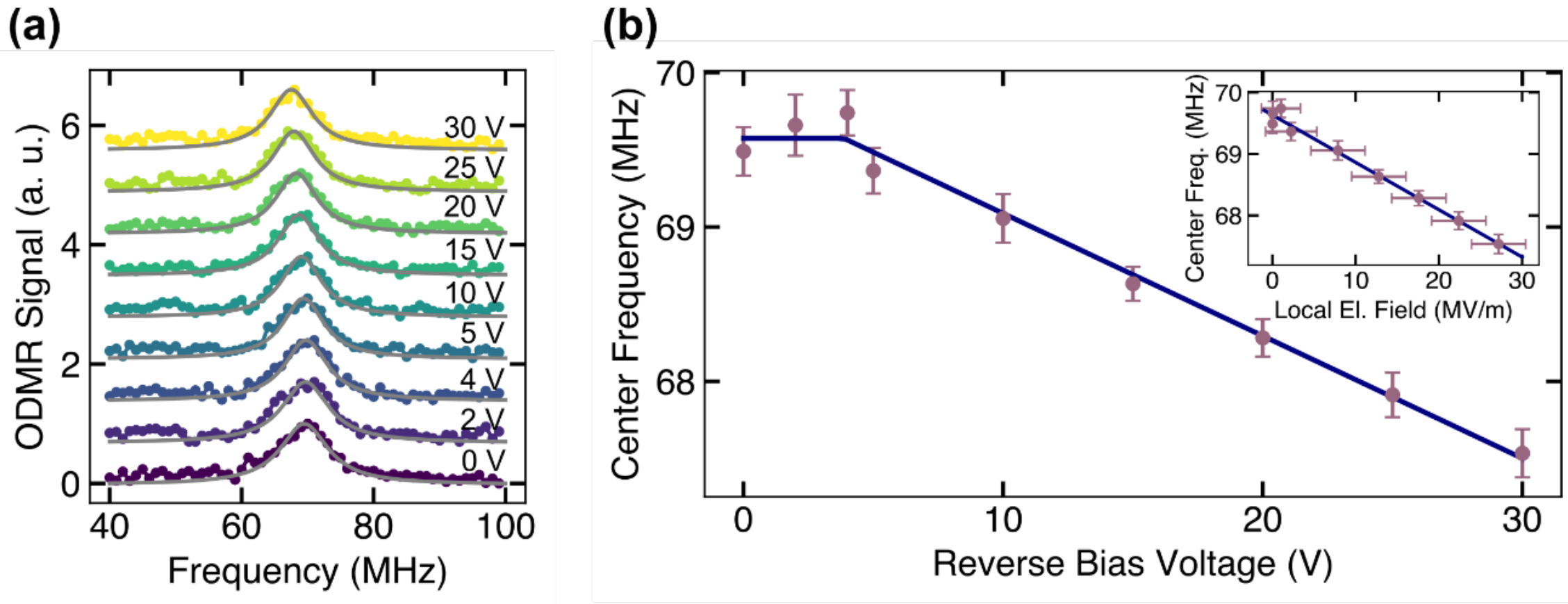

***Figure 4:*** *(a) Offresonant ODMR measurement for different applied reverse bias voltages of 0 V (blue) to 30 V (yellow), performed at room temperature. (b) The shift of the ODMR-peaks reveals that only for voltages greater than* $3.1 \pm 2.4$ *V an effective field at the position of the vacancy forms, shifting the ZFS. Inset: Shift of the ODMR center frequency as a function of the local electric field of* $V_{Si\,2}$ *under the Lorentz local field approximation, revealing a gradient of* $d_{GS} = -79.3 \pm 4.3\ \frac{Hz}{kV/m}$.

To extract the influence of the reverse bias voltage onto the ground state of the $V_{Si\,2}$ center, we performed ODMR measurements at room temperature for reverse bias voltages from $0\ V$ up to $30\ V$ [see Fig. 4 (a)]. There is a clear shift of the resonance position from the ZFS of $\approx 70\ MHz$ visible within the measurements, which is in good agreement with ODMR simulations based on Eq. (2) (see SI10 [32]). This shift is in very good agreement with resonant ODMR measurements as shown in SI10 [32]. For this reason, the electric field at the position of $V_{Si2}$ has the same strength at room and cryogenic temperatures. As the ground state of $V_{Si}$ centers has no temperature dependence [26], we exclude the ZFS shifts as shown in Fig. 4 to originate from heating of the device due to the reverse current. The strength of the shift correlates with the applied reverse bias voltage and a gradient of $d_{GS} = -79.3 \pm 4.3\ \frac{Hz}{kV/m}$. The electric field across the *pin*-diode is oriented from the $n^{++}$ to the $p^{++}$-side, which is parallel to the quantization axis of the $V_{Si}$ center. Our measurements therefore allow not only to infer the absolute value of the electric field, but also whether it is aligned in parallel with the c-axis (0001-direction) or rotated by 180° with respect to the c-axis (000-1 – direction). The onset of the Stark shift, which has been seen in Fig. 2 (b) at around $\sim 2.6\ V$, is also present in these ODMR measurements [Fig. 4 (b)]. This method therefore allows to extract the expansion of the depletion region based on the applied reverse bias voltage at ambient temperature.

In conclusion, we have shown the local-measurement-based reconstruction of physical parameters in an electronic device, like the formation of the space charge region, the electric field, and the doping level with $V_{Si}$ centers in 4H-SiC. The presented method allows the measurement of device characteristics with an Abbe-limited lateral spatial resolution of $\approx 500\ nm$, even far below the surface and at various positions within the semiconductor device. Silicon vacancy centers are also excellent sensors of e.g., temperature [25,59] and magnetic fields [24,60], where the latter can be used to infer the current density in operating integrated circuits [61] with operation regimes from cryogenic up to very high temperatures [62]. We envisage that $V_{Si}$ centers enable the full operando monitoring of semiconductor devices, which can greatly help to infer the device performance characteristics and improve their performance in power electronics. Besides this, the presented method can also contribute to a better understanding of charge carrier concentrations at low temperature.

Besides applications in the electrometry of semiconductor devices, the method presented here is also of great use to establish silicon-vacancy based quantum communication networks. These protocols rely on indistinguishable photons emitted from separate quantum network nodes. This can be achieved by utilizing the presented *pin*-diodes, whose resonances are tuned such that the emitted photons have the same wavelength and two-photon interference [63,64] can occur, which is a prerequisite for spin-spin entanglement [65,66].

**Acknowledgement**

We thank M. Bockstedte (University Linz) for helpful discussions about the Stark shift. R. Nagy acknowledges the funding of BMBF QMNDQCNet (Grant No. 13N16264), DFG (Project No. 462567933), and Lighthouse Project QuMeCo (State of Bavaria). P. Udvarhelyi and P. Narang gratefully acknowledge support by the U.S. Department of Energy, Office of Science, Basic Energy Sciences (BES), Materials Sciences and Engineering Division under FWP ERKCK47 'Understanding and Controlling Entangled and Correlated Quantum States in Confined Solid-state Systems Created via Atomic Scale Manipulation'. This research used resources of the National Energy Research Scientific Computing Center, a DOE Office of Science User Facility supported by the Office of Science of the U.S. Department of Energy under Contract No. DE-AC02-05CH11231 and Award No. BES-ERCAP0029123.

**Author Contributions**

D. S. fabricated the pin-sample and carried out the IV measurements. F. H. performed the PLE and ODMR measurements. F. H. and M. H. performed the sensitivity measurement. J. S. performed the FEM simulations of the pin-diode and carried out the CV measurement. W. K. irradiated the sample. P. U. performed the simulations of the dipole moment and polarizability. H. W. supported sample manufacturing. M. H. performed the data analysis, the ODMR and PLE simulations, wrote the manuscript and coordinated the project. R. N. supervised the project. All authors discussed and contributed to the manuscript.

**Competing interests**

The authors declare no competing interests.

**Data availability**

The data that support the findings of this paper are available from the corresponding author upon reasonable request.

# Supplementary Information - Quantum enhanced electric field mapping within semiconductor devices

## SI1. Sample

We use a commercially available 4H-SiC wafer from JXT Technology Co., Ltd., where according to the manufacturer, on top of a highly n-type substrate (thickness $350\ \mu m$), a $n^{++}$ buffer layer (doping $1 \times 10^{18}\ cm^{-3}$, thickness $1\ \mu m$), an $n^{-}$-type epitaxial layer (doping $2 \times 10^{14}\ cm^{-3}$, according to manufacturer) of $4.1\ \mu m$ thickness (further referred to as "intrinsic layer" of the pin-diode) and a $2\ \mu m$ thick $p^{++}$-epilayer with doping of $2 \times 10^{19} cm^{-3}$ have been grown (see Fig. (SI) 1). At the $n^{++}$- and $p^{++}$-sides of the sample, Ohmic contacts were formed by Ni/Al deposition and subsequent annealing at, respectively, $1000\ ^{\circ}C$ for 5 minutes at the $n^{++}$-side and $800\ ^{\circ}C$ for 3 minutes at the $p^{++}$-side. For the $n^{++}$ contact a layer of 50 nm Ni was used and for the $p^{++}$ contact a composition of 53 nm Ni and 43 nm Al. To create single color centers, the sample was electron irradiated with a flux of $1 \times 10^{12}\ cm^{-2}$ and an energy of 4 MeV, followed by annealing at $600\ ^{\circ}C$ for 30 minutes. As typical defect concentrations in SiC samples with color centers created by electron irradiation are below mid $10^{12}$ $cm^{-3}$ [1] , which is comparable to typical concentrations of defect centers in 4H SiC devices [2], we do not expect any changes of the device characteristics due to electron irradiation.

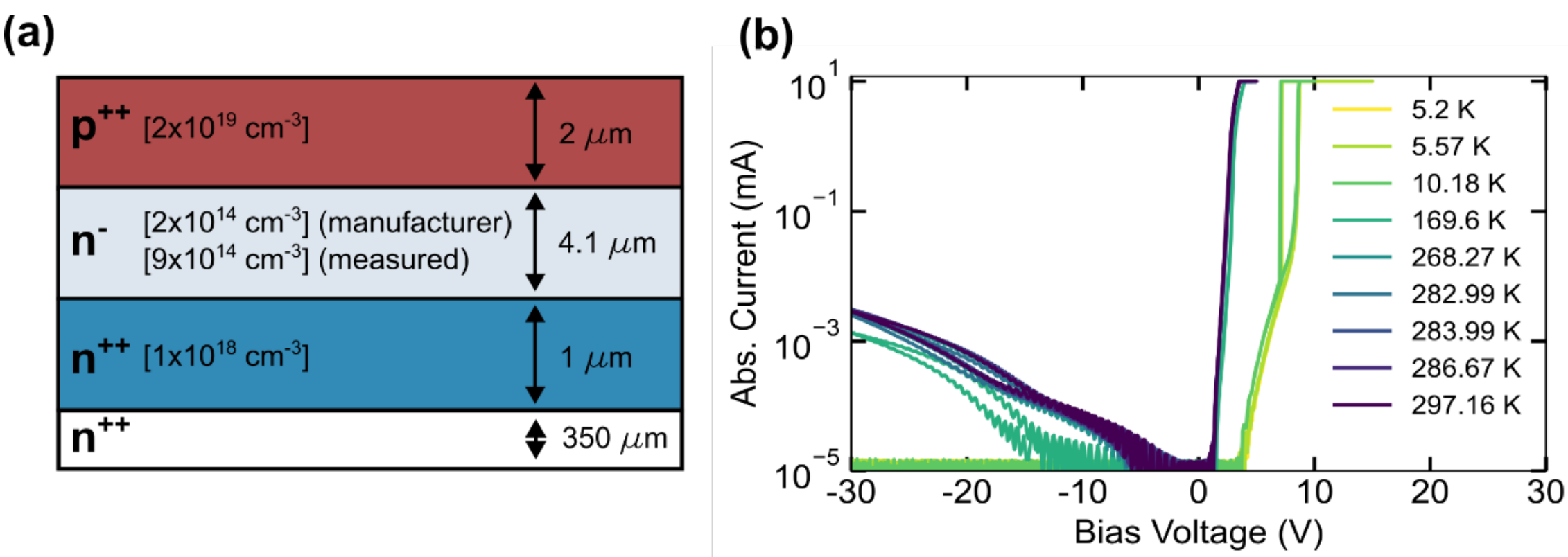

*Figure (SI) 1: (a) Doping layers of the pin-diode. (b) IV measurement of the sample, confirming the characteristics of a pin-diode from 5 K to room temperature. The bias voltage is applied to the $p^{++}$-side of the sample and the $n^{++}$-side being grounded. The diode is operated in forward direction for positive voltages and in reverse bias for negative voltages.*

To confirm the proper manufacturing of our *pin*-diode, we recorded the IV characteristics [see Fig. (SI) 1], where the $n^{++}$-side of the sample was grounded and a voltage was applied to the $p^{++}$-side. Therefore, the diode was operated in forward direction for positive voltages and in reverse bias mode for negative voltages.

From Fig. (SI) 1, one can see that at room temperature, the reverse current increases with reverse bias voltages but takes a value of around $10^{-5}$ mA in reverse direction, which is the minimum current detectable with the used SMU. It can also be seen that the turn-on voltage of the *pin*-diode is shifted towards higher values with decreasing temperatures, which is in accordance with recent measurements of SiC semiconductor devices at cryogenic temperatures [3]. As the IV characteristics of the pin-diode do not differ significantly at room temperature and 5 K, we do not estimate charge carrier concentrations to differ by orders of magnitude.

Besides a bigger intrinsic region (see discussion in main text), the reverse current at room temperature can be reduced in further studies through a structured $p^{++}$-layer. Due to the sample being sawn and cleaved from the waver, edge currents can form which give rise to the relatively large current in reverse bias operation. These edge currents can be reduced via mesa etching on the $p^{++}$-side.“

### CV Measurements of the Doping Concentration in the Intrinsic Layer

To measure the concentration of dopants in the intrinsic layer of our *pin*-diode, we used a sample from the same wafer which is located close to the sample that was used for the spin measurements reported in the main text. Starting from the $p^{++}$-side, 4.5 $\mu m$ were etched away, leaving an effective epitaxial intrinsic layer of $1.6\,\mu m$. On top of this layer, a Schottky contact was fabricated for the CV measurements [see Fig. (SI) 2(a)].

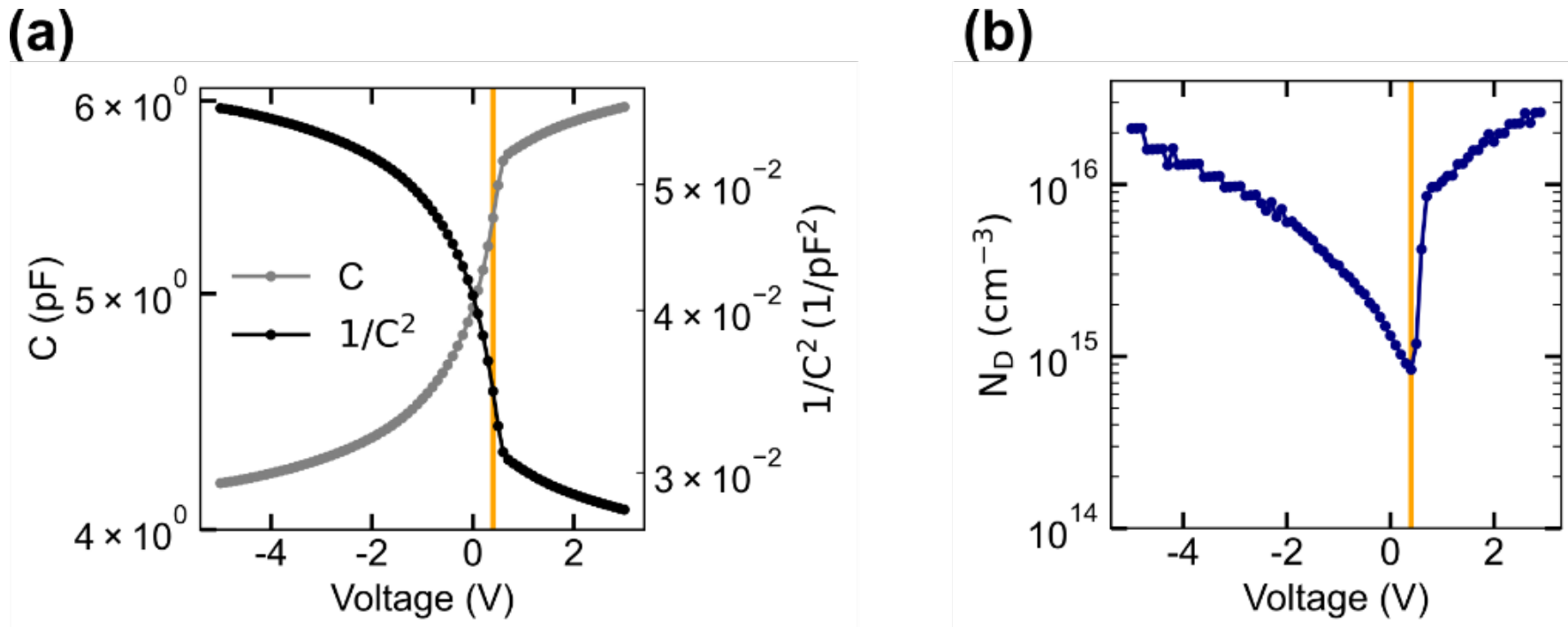

***Figure (SI) 2:*** *(a) Measurements of the capacitance C as a function of the applied voltage (V) and the corresponding $1/C^2$ values. (b) Corresponding concentration of dopants in the intrinsic layer as a function of the applied voltage. Positive voltages in (a) and (b) are in forward, negative voltages in reverse bias direction of the diode.*

The CV measurements were carried out by an Agilent E4980A parameter analyzer with the voltages set around the built-in voltage of the Schottky junction. From the measured capacitance the $1/C^2$ values were determined which are inversely proportional to the doping concentration in the sample, as

$$N_D = \frac{-2}{e\epsilon_0\epsilon_r A^2 \frac{d(1/C^2)}{dV}},$$

with $e$ the electron charge, $\epsilon_r = 10.03$ the relative permittivity of 4H-SiC, $\epsilon_0$ the vacuum permittivity, and $A = (300\,\mu m)^2$ the area of the Schottky contact. In Fig. (SI) 2, two separate regions can be seen which are the reverse and forward direction of the Schottky junction. In reverse conditions, the doping of the epitaxial layer can be determined with the above equation. The doping on the surface of the etched back epitaxial layer was determined to be $8.4 \times 10^{14}\ cm^{-3}$ (for 0.4 V, see orange line in Fig. (SI) 2), increasing towards the buffer layer.

### SI2. Experimental Setup

For the measurements reported here, we use a home-built confocal microscope. For off-resonant excitation, a 730 nm laser (Toptica iBeam smart) is used and for resonant excitation, we use a Toptica DL Pro (916 nm). To create pulses of the resonant laser, an Acousto-Optical Modulator (AOM) from Gooch & Housego (AOMO 3200-1113) is used and the excitation light is power-stabilized at discrete time intervals through an amplitude-EOM (Jenoptik, AM905b). For excitation at both wavelengths, a dichroic mirror (Semrock FF925-Di01) reflects the excitation light and transmits the fluorescence of the color center. The excitation light is reflected at a scanning mirror (Mad City Labs, Nano-MTA2X10) and focused on the sample inside the cryostat (Attocube attodry 800) with a high-NA objective (Zeiss Epiplan-Neofluar 100x, NA 0.9). The emitted fluorescence is collected by the same objective and forwarded to the SNSPD after being highpass-filtered at 950 nm (Thorlabs FELH0950). For the orchestration of our pulse sequences, and to perform time-tagging of the detected counts from the SNSPD we use an OPX from Quantum Machines.

**SI3. FEM-Simulations of Local Electric Field and Band Bending**

For the simulations of the electrical properties, the commercial TCAD (vV-2023.12) software was used. The device was first built in the *sprocess* module and subsequently electrically simulated in the *sdevice* module using the Bank/Rose nonlinear solver with ILS method using parameter set 25 optimized for 4H-SiC simulations. All material properties uused are based on the internal TCAD database, which is widely used in scientific and industrial applications. Physical models considered in the simulation include but are not limited to Shockley-Read-Hall, Avalanche and B2B-Tuneling as recombination mechanism, Doping Dependence and High Field Saturation for the mobility of the charges and Bandgap Narrowing due to doping of the semiconductor.

The electric fields as simulated with TCAD describe the macroscopic field inside the SiC semiconductor structure. As the $V_{Si}$ is a microscopic point defect, it is not only subject to the polarization of the medium under an applied reverse bias voltage but also to electric field contributions stemming from nearby polarized dipoles, which implies that the total electric field acting on the vacancy is [4]

$$E_{tot} = E + E_i,$$

where $E$ is the macroscopic electric field (the external electric field plus the polarization response of the lattice) and $E_i$ is the field generated by the nearby dipoles, which follows to be [4,5]

$$E_i = \frac{P}{3\epsilon_0}.$$

As the polarization $P$ can be expressed in terms of the electric field $E$ via the vacuum permittivity $\epsilon_0$ and the relative permittivity $\epsilon_r$ as $P = \epsilon_0(\epsilon_r - 1)E$, one obtains

$$E_{tot} = E + \frac{P}{3\epsilon_0} = \frac{2 + \epsilon_r}{3} E.$$

With $\epsilon_r = 10.03$ for 4H-SiC [6], one sees that $E_{tot} = 4.01E$, meaning that the actual electric field acting on the $V_{Si}$ is four times higher than the macroscopic field obtained from the TCAD simulations. For the electric field values underlying our findings, we obtain the voltage-dependent electric field by considering the position of the respective $V_{Si}$ center inside the intrinsic layer as identified by the confocal scan (Fig. 1(b), main text) and take the numerically obtained electric field value at this point, multiplied with the correction factor due to the Lorentz local field approximation.

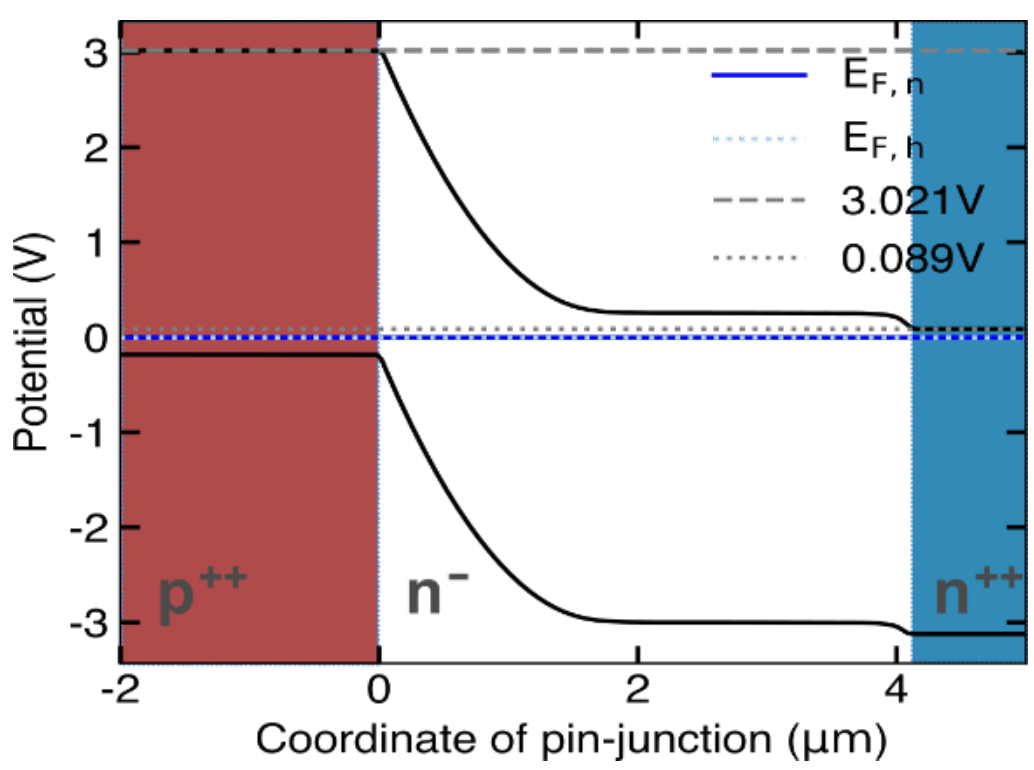

***Figure (SI) 3:*** *Simulated band structure of the pin-diode considered in this work. The n⁻-doped layer of $4.1\mu m$ width and doping of $10^{15}\ cm^{-3}$, in the main text denoted as 'intrinsic layer', is separated by a p⁺⁺ and n⁺⁺ region.*

To obtain the built-in voltage of the *pin*-diode, we simulated the energy bands for an intrinsic doping of $1 \times 10^{15}\ cm^{-3}$ [Fig. (SI) 3]. By subtracting the maximum of the conduction band from its minimum, the built-in voltage of $2.93\ V$ can be obtained. We find that this value does not change significantly for a doping of $8 \times 10^{14}\ cm^{-3}$, as considered in the main text.

We would like to stress that at cryogenic temperatures where our PLE measurements were performed, most of the dopants should be frozen out. Nevertheless, it can be seen from Fig. (SI) 1(b) that our device shows diode-like behavior at temperatures down to 5 K. As suggested by a recent publication [3], we attribute this to Mott transitions, giving rise to ionized donors and acceptors but would like to stress, that the underlying physical mechanisms are subject of current research.

**SI4. Formation of Depletion zone under reverse bias voltages**

The width of the depletion zone at the interface of a *p*- and a *n*-doped region into the n-doped area is

$$x_n(V) = \sqrt{\frac{2\epsilon_0\epsilon_r}{e}\frac{N_A}{N_D}\frac{1}{N_A + N_D}(V_{bi} + V)},$$

where $\epsilon_0$ is the vacuum permittivity, $\epsilon_r = 10.03$ the relative permittivity of 4H-SiC, $e$ the elementary charge, $V$ the externally applied voltage across the junction and $V_{bi} = 2.95\ V$ is the built-in voltage for the given junction. $N_A$ ($N_D$) are the concentrations of acceptors (dopants). As for the device considered here the concentration of acceptors greatly exceeds the number of dopants, $N_A \gg N_D$, the above equation can be reduced to

$$x_n(V) \approx \sqrt{\frac{2\epsilon_0\epsilon_r}{e}\frac{1}{N_D}(V_{bi} + V)},$$

which shows that the length of the depletion zone depends on both the voltage being applied across the diode, as well as the doping inside the intrinsic layer.

The above equation can be used to calculate the concentration of dopants in the intrinsic layer. Rearranging for $N_D$ leads to

$$N_D(V, x_n) = \frac{2\epsilon_0\epsilon_r}{e}\frac{V + V_{bi}}{x_n^2}.$$

$V_{Si\ 2}$ has a threshold voltage of $2.6 \pm 0.4\ V$ [see inset Fig. 2(b), main text] and its distance to the interface between $p^{++}$ and intrinsic layer is $2.71 \pm 0.25\ \mu m$. Note that the uncertainty of the position considers that the resolution of our confocal setup is on the order of 500 nm. With this we find the concentration to vary from $6.52 \times 10^{14}\ cm^{-3}$ $(V = 2.2\ V, x_n = 2.71 + 0.25\ \mu m)$ to $10.90 \times 10^{14}\ cm^{-3}$ $(V = 3\ V, x_n = 2.71 - 0.25\ \mu m)$, resulting in a doping concentration of $N_D = (8.7 \pm 2.2) \times 10^{14}\ cm^{-3}$.

### SI5. Additional Information about PLE-Measurements

For the underlying PLE sequence of the measurements of the Stark Shift [Fig. 2(b), main text], an initial red laser pulse at 730 nm ensures the correct charge state. Then the wavelength of the diode laser (Toptica DL Pro) is tuned through an external DC voltage. While sweeping, the wavelength of the laser is measured by a wavemeter (High Finesse WS7), and the counts are detected with an SNSPD (Single Quantum). For higher spin contrast, a microwave AC voltage at 70 MHz is applied through the copper wire in the vicinity of the color centers [see Fig. 1(b), main text]. Both resonances [see Fig. 1(d) for an example] appear as peaks in the obtained spectrum. In Fig. (SI) 4, the detuning between both resonances is plotted, where no significant deviation from $\Delta_{A_1,A_2} \approx 1\ GHz$ over the applied voltages can be found.

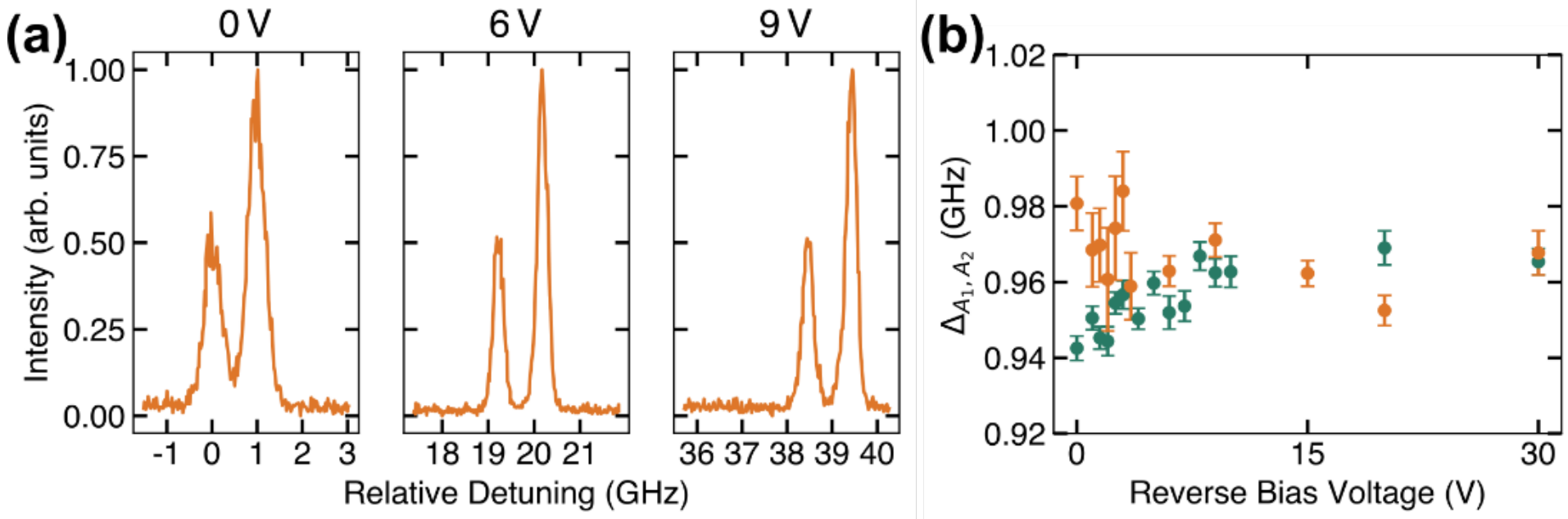

***Figure (SI) 4:*** *(a) Examples of PLE measurements underlying the Stark shift as shown in Fig. 2(b) (main text) for applied reverse bias voltages of 0V, 6V and 9V for $V_{Si2}$. (b) Detuning between the $A_1$ and $A_2$ transitions of vacancies $V_{Si\ 1}$ (green) and $V_{Si\ 2}$ (orange) for different applied voltages. The Stark shift of the $A_1$ transition is shown in Fig. 2(b) (main text).*

The linewidths [Fig. 3(a), main text] were obtained purely optically with a phase-EOM (Exail MPX-950-LN-10), where the seed laser was detuned by +1 GHz from the $A_2$ transition. After a 730 nm pulse to ensure the correct charge state, the AC voltage applied to the phase-EOM was modulated from 800 MHz to 1.4 GHz, which effectively sweeps the sidebands over the $A_2$ transitions. The signal was fitted with a Lorentzian to obtain the FWHM as shown in Fig. 3(a).

### SI6. Dipole Moments and Polarizabilities for Different Doping Concentrations

As the concentration of donors inside the intrinsic layer of the sample used for the measurements reported here has some uncertainty concerning their concentration (see CV-measurements in SI1 as well as discussion of the relation between the Stark shift threshold and the concentration in SI4), we simulated the electric field across our *pin*-diode as a function of the applied voltage for concentrations of $7 \times 10^{14}\ cm^{-3}$, $9 \times 10^{14}\ cm^{-3}$ and $11 \times 10^{14}\ cm^{-3}$, which covers the full range of the concentration of dopants as derived in SI4.

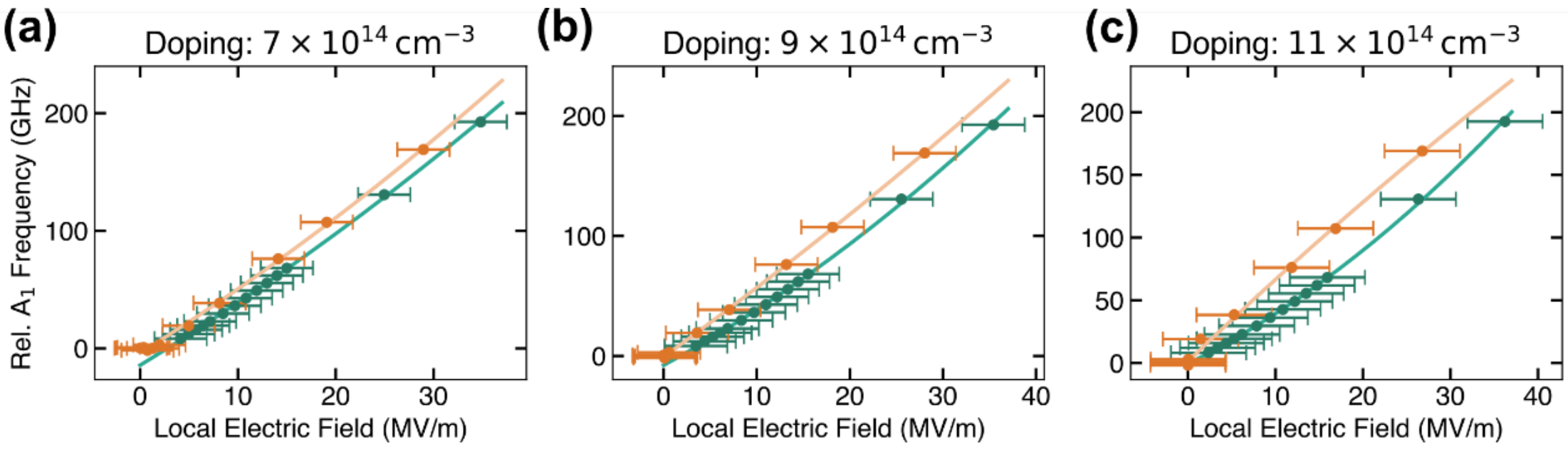

***Figure (SI) 5:*** *Stark shift of $V_{Si\ 1,2}$ obtained from Fig. 2(b) (main text) by converting the applied voltage into the local electric field at the position of the $V_{Si}$ center and by considering a doping of (a) $7 \times 10^{14} cm^{-3}$, (b) $9 \times 10^{14} cm^{-3}$ or (c) $11 \times 10^{14} cm^{-3}$ in the simulations to cover the error of the estimated doping as given in the main text.*

As one can see in Fig. (SI) 5, the Stark shift as a function of the local electric field behaves slightly differently, depending on the concentration of dopants in the intrinsic layer considered in the FEM simulations.

To extract the dipole moment $d$ and the polarizability $\alpha$, the relative $A_1$ frequency $\Delta f_{A_1}$ is fitted by

$$\Delta f_{A_1} = -d \cdot E - \frac{\alpha}{2} E^2 + f_0. \qquad \text{(SI 1)}$$

The resulting fit parameters are listed in Tab. (SI) 1. The dipole moment is slightly higher for $V_{Si\ 2}$, compared to $V_{Si\ 1}$, which we attribute to effects not covered by our electric field simulation, like e.g., screening effects leading to local inhomogeneities of the electric field.

| **(a)** | $N_D = 7 \times 10^{14}\ cm^{-3}$ | | |
|---|---|---|---|
| | $d \left(\frac{GHz}{MV/m}\right)$ | $\alpha \left(\frac{GHz}{\left(\frac{MV}{m}\right)^2}\right)$ | $f_0\ (GHz)$ |
| $V_{Si\ 1}$ | $-5.07 \pm 0.14$ | $-0.05 \pm 0.01$ | $-14.35 \pm 1.00$ |
| $V_{Si\ 2}$ | $-5.14 \pm 0.41$ | $-0.06 \pm 0.03$ | $-3.90 \pm 1.38$ |

| **(b)** | $N_D = 9 \times 10^{14}\ cm^{-3}$ | | |
|---|---|---|---|
| | $d \left(\frac{GHz}{MV/m}\right)$ | $\alpha \left(\frac{GHz}{\left(\frac{MV}{m}\right)^2}\right)$ | $f_0\ (GHz)$ |
| $V_{Si\ 1}$ | $-4.20 \pm 0.15$ | $-0.09 \pm 0.01$ | $-7.73 \pm 0.96$ |
| $V_{Si\ 2}$ | $-5.59 \pm 0.13$ | $-0.03 \pm 0.01$ | $-0.67 \pm 0.42$ |

| **(c)** | $N_D = 11 \times 10^{14}\ cm^{-3}$ | | |
|---|---|---|---|
| | $d \left(\frac{GHz}{MV/m}\right)$ | $\alpha \left(\frac{GHz}{\left(\frac{MV}{m}\right)^2}\right)$ | $f_0\ (GHz)$ |
| $V_{Si\ 1}$ | $-3.45 \pm 0.10$ | $-0.11 \pm 0.01$ | $-0.49 \pm 0.61$ |
| $V_{Si\ 2}$ | $-6.69 \pm 0.39$ | $0.03 \pm 0.03$ | $1.46 \pm 1.36$ |

***Table (SI) 1:*** *Dipole moments $d$ and polarizabilities $\alpha$ extracted via the fit of $\Delta f_{A_1}$ from the Stark shift measurements shown in Fig. (SI) 3 for different considered doping concentrations in the intrinsic layer.*

**SI7. First-principles calculations of dipole moment and polarizability**

We apply first-principles calculations to accurately characterize the changes in the dipole moment and polarizability during the excitation process in the $k$-$V_{Si}$ defect. As opposed to the method in Ref. [7], we use the modern theory of polarization [8] to extract these properties from bulk calculations instead of the simulation of a hydrogen-terminated thin surface model. Furthermore, we obtain these values directly from the calculated macroscopic dipole moment and its variation, as well as from the calculated ZPL shift in the presence of a local electric field in the crystal. The former method is expected to estimate the properties of the $V_2$ center more accurately, as the calculated ZPL energy suffers from the bandgap problem common in local functional calculations. We apply density functional theory (DFT) calculations using the PBE functional [9] as implemented in Quantum Espresso [10–12]. Plane-wave and charge density cutoffs of 80 and 320 Ry were used, respectively. Core electrons were treated using Optimized Norm-Conserving Vanderbilt (ONCV) pseudopotentials [13,14]. The defect is modeled in a large 576-atom supercell model of bulk 4H-SiC. Owing to the Brillouin zone folding, we sample only the Gamma point of the reciprocal space. The ground and first excited state of the $V_2$ center is calculated in the ΔSCF method [15] with the relaxation of the atomic positions. A local electric field is applied along the symmetry axis of the defect (c-axis) reflecting the experimental setup. Thus, we calculate the change in the axial component (z) of the dipole moment vector ($\Delta d_z = d$) and the polarizability matrix ($\Delta \alpha_{zz} = \alpha$) in the excitation process. The piezo effect is not calculated, i.e., the optimized ground and excited state geometries are not relaxed further in the presence of the electric field. Our results [see Fig. (SI) 6] show a linear electric field dependence of $d$, suggesting that the second-order expansion of the Stark shift in Eq. (3) accurately describes the measured ZPL energy shift. From the linear fit to our calculation data and its standard error, we determine $d = 1.83 \pm 0.01 \, \frac{GHz}{\left(\frac{MV}{m}\right)}$ and $\alpha = -0.0254 \pm 0.0006 \, \frac{GHz}{\left(\frac{MV}{m}\right)^2}$. The differences from our measured values are probably related to the functional used in the calculation and to piezo effects. As these improvements in the calculation method are orders of magnitude more expensive, they are beyond the scope of our current study.

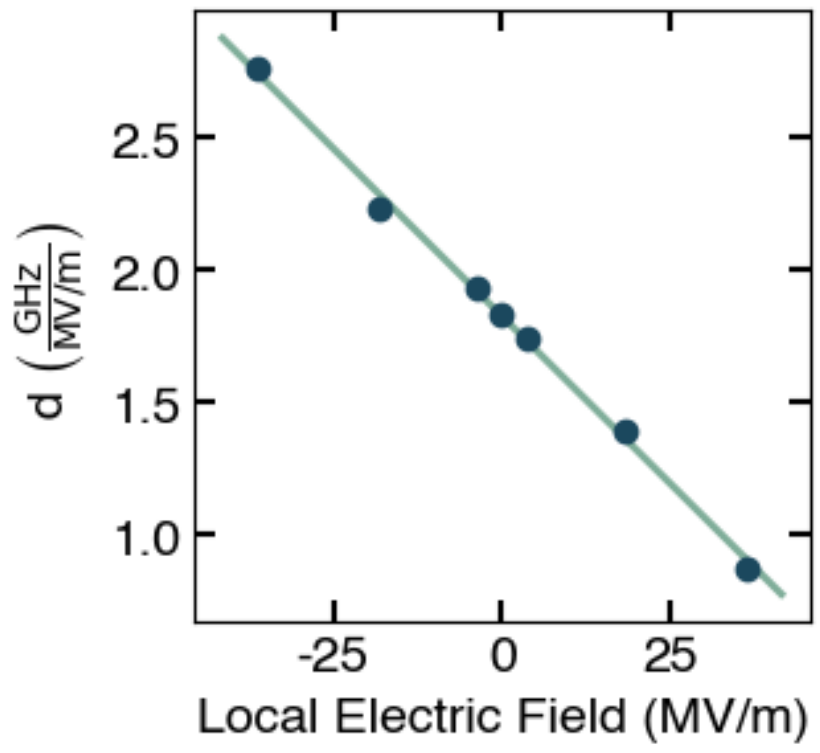

***Figure (SI) 6:*** *Simulation of the electric field dependence on the dipole moment.*

## SI8. Sensitivity of Stark Shift Measurements

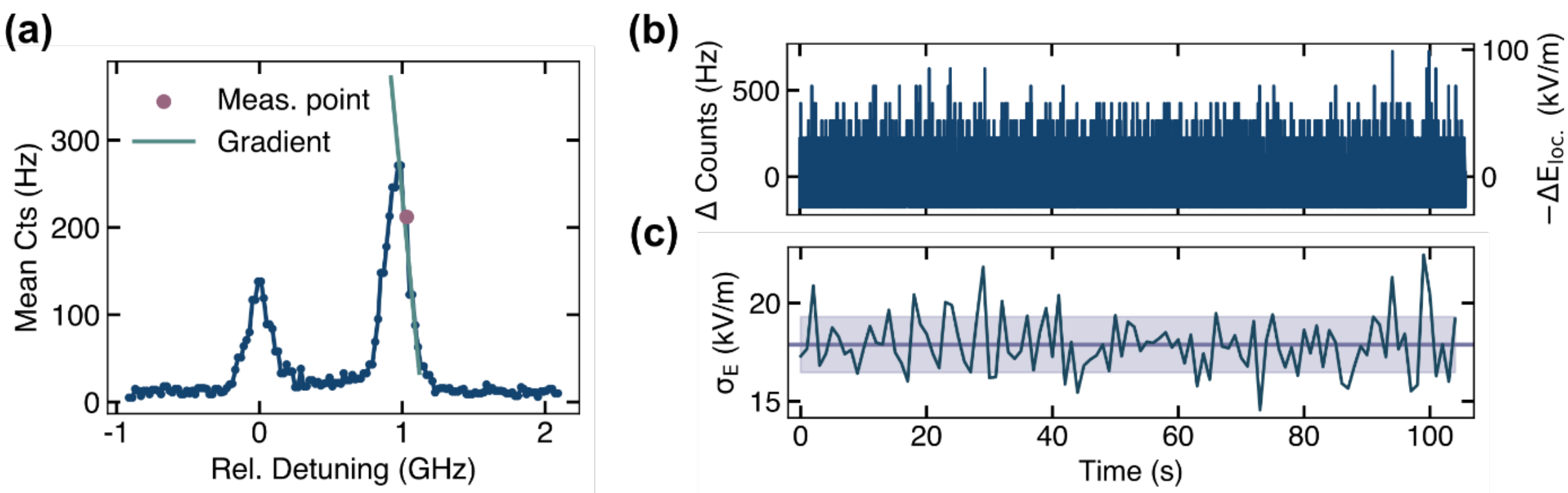

***Figure (SI) 7**: (a) To infer the sensitivity of the Stark shift measurements of Fig. 2(b,c) (main text), a PLE measurement at $V_{Si\,2}$ was performed with 30 V applied to the diode in reverse bias. The red point represents the maximum gradient, having the highest sensitivity. (b) Counts converted into electric field fluctuations. (c) Standard deviation of (b) in 1s intervals.*

To infer the sensitivity of the Stark shift measurements shown in Fig. 2(b,c), we performed a PLE measurement [Fig. (SI) 7(a)] at 30 V applied to $V_{Si\,2}$ (see main text). At the maximum of its gradient (red dot in Fig. (SI) 7(a)), a time series was recorded for more than 100 seconds [Fig. (SI) 7(b)] with a sampling rate of 100 Hz.

To convert the measured photoluminescence fluctuations at that point $\Delta cts(t)$ [Fig. (SI) 7 (b)] into electric field fluctuations $\Delta E(t)$, one can use that

$$\Delta \mathrm{E(t)} = \frac{\mathrm{d}E}{\mathrm{d}f} \cdot \frac{\mathrm{d}f}{\mathrm{d}cts}\bigg|_{WP} \Delta cts(t).$$

The gradient at the working point ($WP$, see Fig. (SI) 7 (a)), $\mathrm{grad} = \mathrm{d}cts/\mathrm{d}f$, can be converted into $\mathrm{d}cts/\mathrm{d}f|_{WP} = 1/\mathrm{grad}$ due to the linearity of the PLE around this point. To infer the gradient of the electric field with respect to the frequency, we use that for 30 V one has $d \cdot E \gg \frac{\alpha}{2} E^2, f_0$, such that from Eq. (SI 1) it follows that

$$\frac{\mathrm{d}E}{\mathrm{d}f} = \frac{1}{\mathrm{d}f/\mathrm{d}E} = \frac{1}{-d},$$

which allows to convert the fluctuations of counts into electric field fluctuations [Fig. (SI) 7(b)]. The sensitivity is then given as the mean value of the standard deviations of these electric field fluctuations within 1s intervals [Fig. (SI) 7(c)], which we find to be

$$\eta = 17.89 \pm 1.43 \,\frac{kV/m}{\sqrt{Hz}}.$$

This value is higher than the sensitivity reported by Ref. [16] for Stark-shift measurements of divacancies, but comparable to electric field sensing experiments with single NV-centers in bulk diamond [17].

## SI9. Estimation of Linewidth-Broadening due to Reverse Current

The current as displayed in Fig. 3(b) (main text) was simulated from the number of charge carriers having flown through a cross section of $1\ cm^2$. Their concentration within measurement time $\tau$ can thus be inferred as

$$n_e = \frac{j}{q \cdot v_e},$$

where $j$ is the current density and $q$ the electron charge and $v_e$ the drift velocity of electrons in 4H-SiC, which was simulated to be $v_e \sim 10^7 \frac{cm}{s}$ for elevated applied reverse voltages. Using the current density as displayed in Fig. 3(b) (main text), one obtains an electron density on the order of $\sim 10^8\ cm^{-3}$, as depicted in Fig. (SI) 8(a).

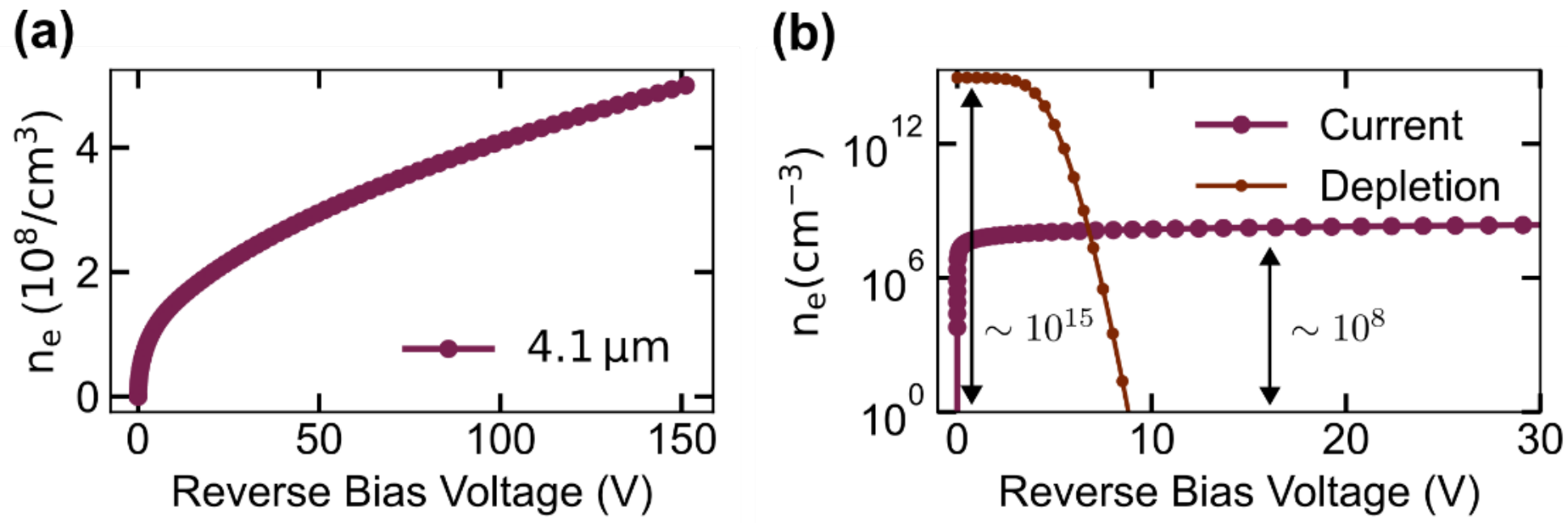

***Figure (SI) 8:*** *(a) Electron density due to the reverse current flowing through the pin-diode. (b) Schematic to illustrate competition between the depletion due to the space charge region and the injection of charge carriers due to the reverse bias current.*

The total amount of charge carriers inside the intrinsic layer of the *pin*-diode is the result of two competing processes, as schematically illustrated in Fig. (SI) 8(b). On the one hand, depletion has the ability to reduce the number of charge carriers by $\sim 10^{15}\ cm^{-3}$ but the reverse bias current causes an injection of electrons on the order of $\sim 10^8\ cm^{-3}$. For this reason, depletion causes an effective reduction by $\sim 10^7\ cm^{-3}$.

The PLE linewidth is governed by the total electric field fluctuations in the vicinity of the $V_{Si}$ center. As this will not scale linearly with the concentration of electrons, we do not expect a linear dependence between the PLE linewidth and the concentration of electrons.

### SI10. Effect of Magnetic Fields on the PLE spectra

Even though there should be no magnetic field inside the *pin*-diode for a homogeneously distributed reverse current, we acknowledge that local inhomogeneities of the Ohmic contacts might cause spatially confined currents, which can cause a magnetic field.

Based on the Lindblad Master equation, we developed a model of the $V_{Si}$ center, which includes the metastable-state dynamics and which can account for magnetic fields in both the ground and excited state. The transition rates in and out of the metastable states as well as the radiative decay rates were taken from D. Liu et al. [18] and the simulations (Fig. (SI) 9) were performed with the open-source Python package QuTiP [19,20].

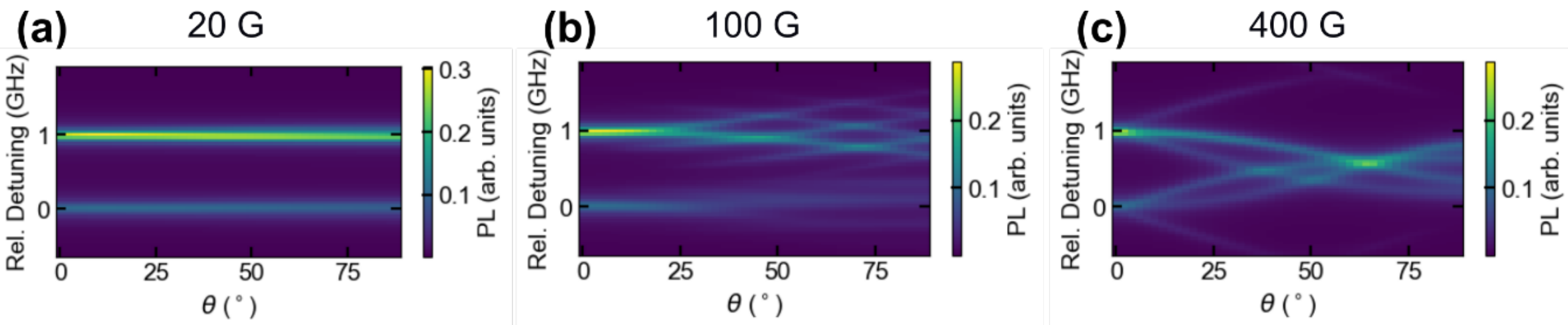

*Figure (SI) 9: Simulated PLE spectra for magnetic field strengths of (a) 20G, (b) 100G and (c) 400G as a function of the altitude angle $\theta$ for an azimuthal angle of $\varphi = 45^{\circ}$.*

In Fig. (SI) 9 one can see the resulting PLE spectra for absolute magnetic field values from 20 G to 400 G as a function of the altitude angle $\theta$ and for an azimuthal angle of $\varphi = 45^{\circ}$. If the magnetic fields are either sufficiently weak (Fig. (SI) 9(a)) or aligned along the c-axis of the $V_{Si}$ center (Fig. (SI) 9(b,c) for $\theta = 0$), the PLE signal is not altered by magnetic fields. Only if the magnetic fields are sufficiently high and possess a sufficient degree of transversal components, the relative splitting between the $A_1$ and $A_2$ peaks is influenced. As the relative distance between $A_1$ and $A_2$ stays constant throughout our measurements (see Fig. (SI) 4(b)), we therefore conclude that magnetic fields generated from reverse currents will not influence the measurements reported here and will only cause linewidth broadenings of the PLE signals as discussed in the main text.

## SI10. ODMR Simulations and resonant ODMR measurements

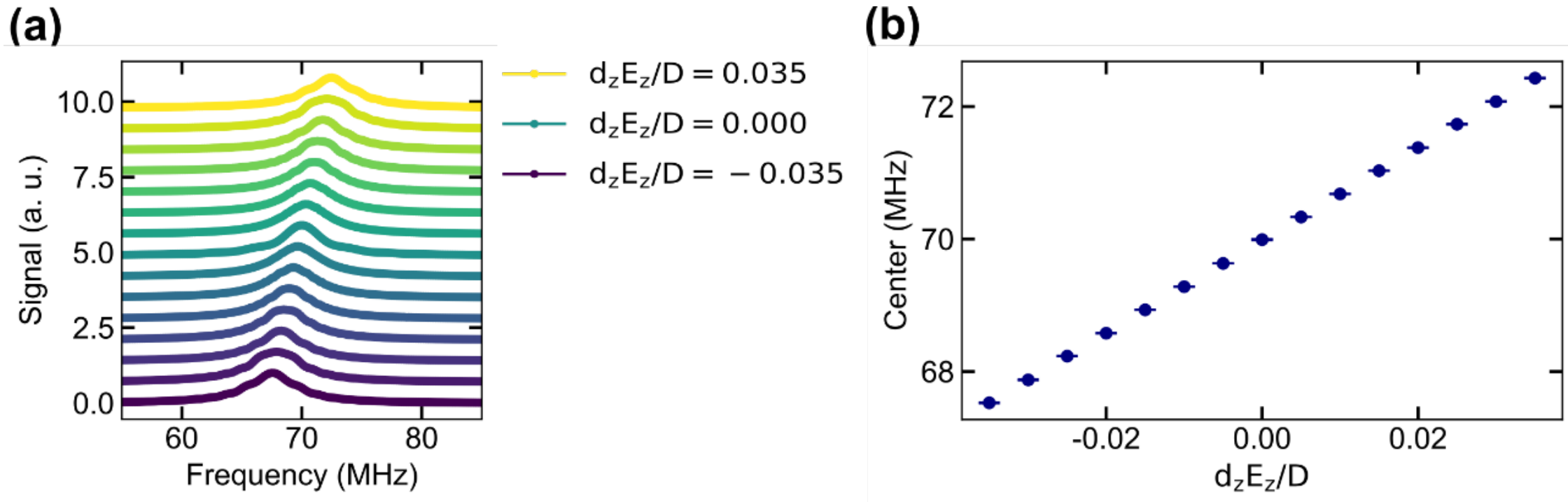

*Figure (SI) 10: (a) Simulated ODMR for $d_z E_z/D$ ranging from -0.035 to 0.035 in steps of $5 \times 10^{-3}$. (b) Extracted center positions of the ODMR peaks from (a), showing linear dependence on the axial electric field $E_z$.*

For a deeper understanding of the ODMR results presented in Fig. 4 (main text), we perform simulations with the open-source Python package QuTiP [19,20] and consider the ground state of the $V_2$ center to be described by [17,21]

$$H_{gs} = (D + d_z E_z) S_z^2.$$

Here $2D = 70\ MHz$ is the $V_2$-center's ZFS, $d_z E_z$ is the electric field component along the c-axis of the lattice. Effects of transverse electric fields $E_{x,y}$ are neglected due to the device geometry. We would like to stress that the above Hamiltonian was derived for NV-centers in diamond. As $V_{Si}$ centers also obey $C_{3v}$ symmetry, we do not expect any changes in the contributions with respect to the ZFS.

After ensuring the correct charge state with a 730 nm laser pulse, we measured resonant ODMR where the frequency of the applied microwave field is varied while the system is read out with a laser pulse resonant with $A_2$. As we start with a microwave frequency which does not cause any transition between $|\pm 1/2\rangle$ and $|\pm 3/2\rangle$, the $A_2$ laser effectively pumps the $V_2$ center into $|\pm 1/2\rangle$, which causes the initial state to be $\rho_{init} = \frac{1}{2}\left[|+\frac{1}{2}\rangle\langle+\frac{1}{2}| + |-\frac{1}{2}\rangle\langle-\frac{1}{2}|\right]$. Only if the microwave drives a transition from $|\pm 1/2\rangle$ to $|\pm 3/2\rangle$ a photon is detected, which is why we calculate the expectation value of the time-evolved state with $\frac{1}{2}\left[|+\frac{3}{2}\rangle\langle+\frac{3}{2}| + |-\frac{3}{2}\rangle\langle-\frac{3}{2}|\right]$. As can be seen from Fig. (SI) 10, the central ODMR peak is shifted towards higher values for positive and reduced from 70 MHz for negative electric fields.

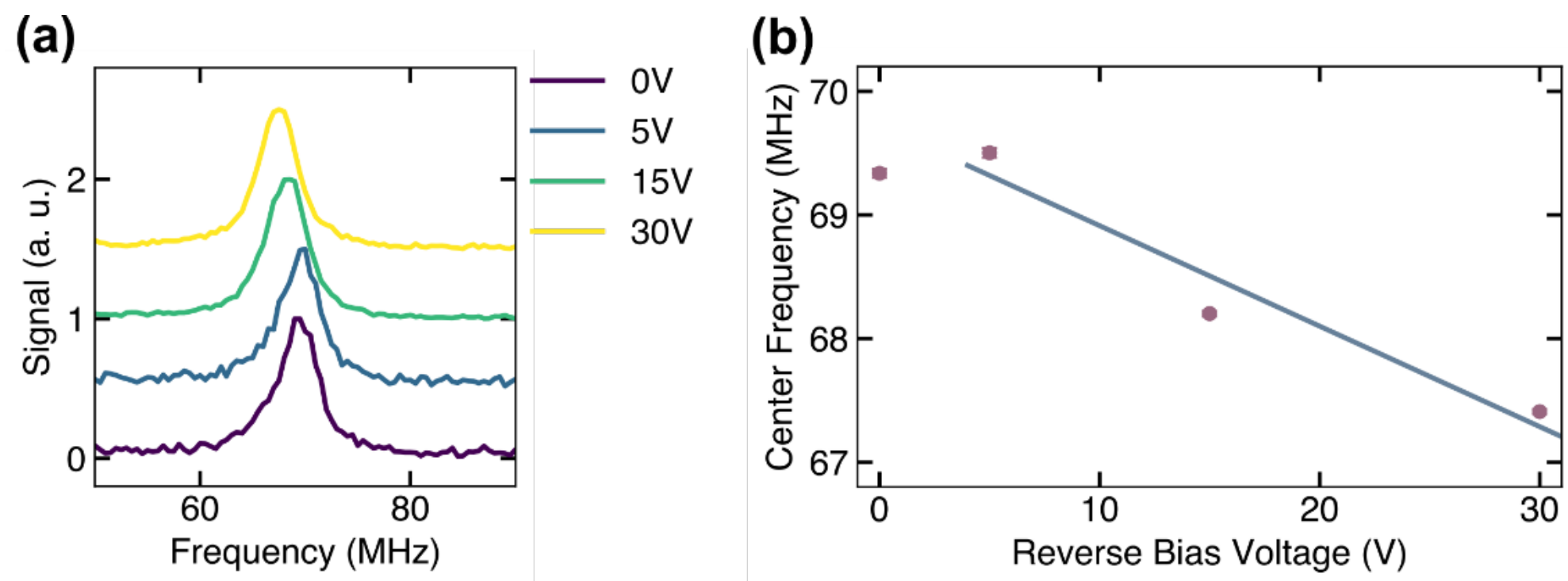

*Figure (SI) 11: (a) Resonant ODMR measurement of $V_{Si2}$ for different applied reverse bias voltages of 0 V (blue) to 30 V (yellow), performed at 5 K. (b) The shift of the ODMR-peaks reveals that only for voltages greater than 5 V an effective field at the position of the vacancy forms, shifting the ZFS.*

Besides the offresonant ODMR measurements at room temperature as reported in the main text, we also performed resonant ODMR measurements at 5 K with the 916nm laser resonant with the $A_2$ transition (see Fig. (SI) 11), revealing that only for reverse bias voltages $> 5\,V$ the effective ZFS is shifted towards lower frequencies. From Fig. (SI) 10(b) one can see that therefore $d_z E_z < 0$. As the electric field is oriented parallel to the c-axis of the 4H-SiC lattice (and not anti-parallel), this indicates that $d_z < 0$. From a linear fit of reverse bias voltages > 5 V, we obtain $d_{GS} = -77.9 \pm 12.0\ \frac{Hz}{kV/m}$, which is in good agreement with the value obtained from the offresonant ODMR measurements from the main text.

## SI11. Charge states of silicon vacancy centers for applied reverse bias voltages

Throughout our experiments we did not observe any change of the charge state of the measured silicon vacancy centers due to applied reverse bias voltages. As shown by Ref. [22], the transition of a k-$V_{Si}$ center from its neutral to the single negative charge state happens at energies of 1.29 eV above the valence band and from negative to double-negative for 2.59 eV. Based on TCAD simulations of the band structure, we find that for 0 V (Fig. (SI) 12(a)) a considerable region from the $n^{++}$-side into the intrinsic layer has a Fermi energy higher than the $V_{Si}^{-} \leftrightarrow V_{Si}^{2-}$, making the $V_{Si}$ center to be in its double negative charge state. Already for 5 V (see Fig. (SI) 12(b)), the region where the Fermi levels lie above the $V_{Si}^{-} \leftrightarrow V_{Si}^{2-}$ transition is shifted more towards the $n^{++}$ layer. In Fig. (SI) 12(c,d) we simulated, respectively, the (quasi-) Fermi levels for $V_{Si1}$ and $V_{Si2,3}$ which shows that for all reverse bias voltages >0 V the silicon vacancy centers are found to be in their single negative charge state. This observation is in accordance with recent findings [23], where no change in the power-dependent ionization rate of *k*-$V_{Si}$ centers depending on the applied reverse bias voltage in a Schottky diode was found.

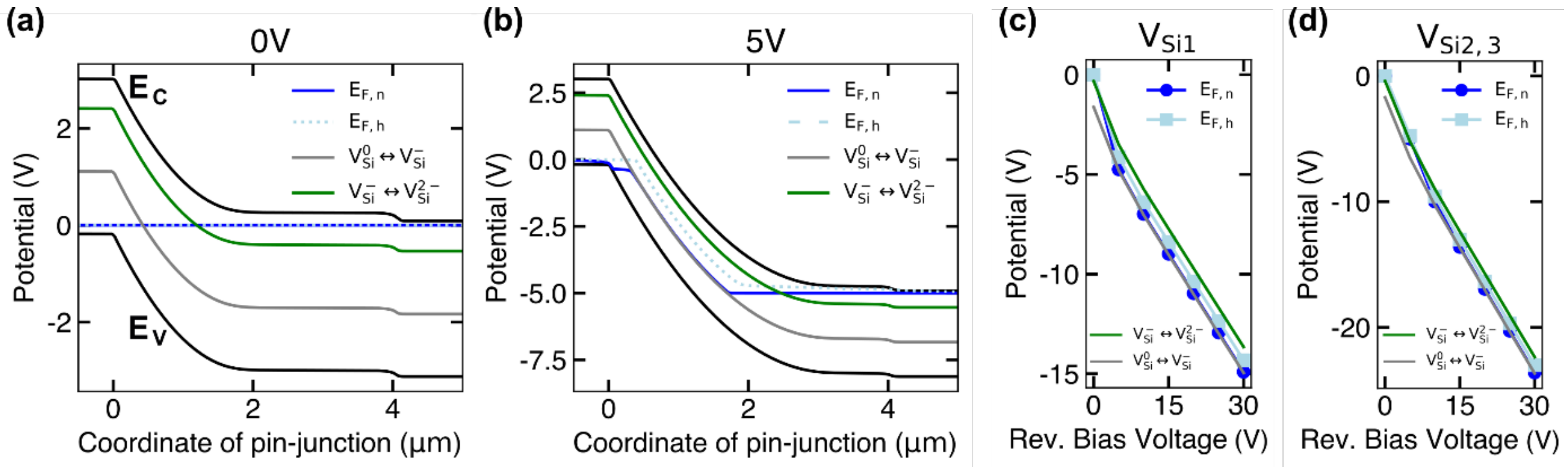

***Figure (SI) 12:*** *Band structure of our pin-diode of (a) 0 V and (b) 5 V reverse bias voltages, including the transition energies of of $V_{Si}^0 \leftrightarrow V_{Si}^-$ and $V_{Si}^- \leftrightarrow V_{Si}^{2-}$. The quasi-Fermi energies for electrons and holes, $E_{F,n}$ and $E_{F,h}$ of (c) $V_{Si\,1}$ and (d) $V_{Si\,2,3}$ stay between the charge state transitions of $V_{Si}^0 \leftrightarrow V_{Si}^-$ and $V_{Si}^- \leftrightarrow V_{Si}^{2-}$ for bias voltages >0 V, which causes the investigated silicon vacancy centers to stay in the single negatively charge state throughout our measurements.*